\documentstyle[prd,aps,preprint,axodraw]{revtex}
\tightenlines
\input{epsf.sty}

\newcommand{\be}{\begin{equation}}
\newcommand{\ee}{\end{equation}}
\newcommand{\bea}{\begin{eqnarray}}
\newcommand{\beas}{\begin{eqnarray*}}
\newcommand{\eea}{\end{eqnarray}}
\newcommand{\eeas}{\end{eqnarray*}} 
\newcommand{\ba}{\begin{array}}
\newcommand{\ea}{\end{array}}

\begin{document}

\draft
\preprint{\vbox{\hbox{UMD-PP-01-047}}}

\bigskip
\bigskip

\title{ Geometric CP Violation with Extra Dimensions}
\author{Darwin Chang$^{1,3}$\footnote{e-mail: chang@phys.nthu.edu.tw} and
R. N. Mohapatra$^2$\footnote{e-mail:rmohapat@physics.umd.edu},\\
$^1$ Physics Department, National Tsing-Hua University, Hsinchu, Taiwan,
30043, ROC\\
$^2$ Department of
Physics, University of Maryland, College Park, MD, 20742\\
$^3$ Physics Department, University of Illinois at Chicago, Chicago, IL,
60607-7059}

\date{March, 2001}

\maketitle
\begin{abstract}
{We discuss how CP symmetry can be broken geometrically through orbifold
construction in hidden extra dimensions in the context of D-brane
models for  particle 
unifications. We present a few toy models to illustrate the idea and suggest
ways to incorporate this technique in the context of realistic models.}
\end{abstract}
\bigskip
\bigskip

\section{Introduction}
Origin of discrete symmetry violations observed in nature, such as 
parity or CP violations, is still a mystery. In the standard model of
Glashow, Weinberg and Salam, parity violation was put in ``by hand''
by excluding the right handed neutrinos from the theory.
Similarly in the three generation Kobayashi-Maskawa extension, the lone
CP violating phase is also an input that arose by making the Yukawa couplings
complex and no insight is gained as to how nature broke CP
symmetry\cite{book}.

In early 1970's, it was pointed out that if one assumed the standard model 
to be a part of the left-right symmetric model of weak interactions, a
more satisfactory framework for the origin of parity violation can be 
obtained\cite{lr}. The recent discovery of neutrino masses and their
understanding in terms of the seesaw mechanism may be construed as a sign
pointing towards ultimate left-right symmetry of Nature.
The question one must then address is how left-right
symmetry is broken in Nature.  

Ever since Einstein's general theory of relativity, physicist have often
tried to see if an idea or a theory can be realized
geometrically.  The door to this possibility has been open more widely by
the realization that the Nature can accommodate extra dimensions of
spacetime as long as they are sufficiently hidden. This has received
strong theoretical support from superstring theories which require the
existence of extra hidden dimensions for their consistency.
In this context, an intriguing mechanism for explaining the existence of
broken symmetries is to connect them with the existence of hidden
compact dimensions in Nature. Some examples realizing this for parity were
suggested in Ref. \cite{perez}. In fact it is tempting to speculate,
whether all symmetry violations in Nature owe their origin to the presence
of extra dimensions. Recently, it has been suggested that weak gauge
symmetry \cite{hall} as well as the grand unification symmetries
\cite{kawamura} could be broken by the extra dimensional effects.

In this brief note, we explore the possibility that CP symmetry breaking
is connected with the existence of hidden compact dimensions in Nature. We
present several simple examples where the CP symmetry is broken by
orbifold projections. We then suggest ways to incorporate them
into more realistic models.

\section{A toy model}
 
Consider a simple $U(1)$ gauge model with four left-handed fermions,
$f_{1,3}$ with
 charge +1 and $f_{2,4}$ with charge -1 and one real scalar $\eta$ which
is $U(1)$ neutral. We 
will use flat fifth dimension so that when the fifth dimension $y$ is
compactified, we can use sine and cosine expansion. All
the fermion fields are  assumed to be bulk fields (with four 4-dimensional
chiral fermions combined into two 5-dimensional fermions) and the
$\eta$ field is assumed to be located in the brane at $y=0$. 
Let us assume that the fifth dimensional space is projective and
therefore the Lagrangian is required to be invariant under $y\rightarrow
-y$ tranformation and all the bulk fields tranformation under the even or
odd representation of an asscoated $Z_2$ symmetry.
We shall
assume the brane fields (including $\eta$) to be even under $Z_2$. As far
as the bulk fermions go, invariance of the bulk kinetic energy term
 $\bar{f}_{iL}\partial_y f_{iR}$ under $Z_2$ implies that $f_{iL}$ and
$f_{iR}$ have opposite $Z_2$. This means that if one of them is odd under
$Z_2$, the other is even. Let us assume that  $f_{1L},f_{2L}$ are 
$Z_2$ even while $f_{3L}, f_{4L}$ are $Z_2$ odd.

 The 5-dimensional Lagrangian for this model relevant for our discussion 
can be written as:

\begin{eqnarray}
{\cal L}(x,y)= \frac{1}{{M_*}}\delta(y)\left[\lambda \eta
(f^T_{1L}C^{-1}f_{2L} - (f_{3L})^{cT}C^{-1} (f_{4L})^c)
\right]  \nonumber\\
+\mu (f^{T}_{1L}C^{-1} f_{2L} + (f_{3L})^{cT}C^{-1} (f_{4L})^c) 
+ H.c.
\end{eqnarray}
where $(f_L)^c$ is the usual 4-dimensional charge conjugation of $f_L$.
The above coupling structure can obviously be maintained naturally
through some flavor symmetry ($U(1)\times U(1)$ for example) for the
fermions $f_i$ which we shall not elaborate.
Under CP symmetry, we define the fields to transform as: 

\begin{eqnarray}
f_{1L} \rightarrow (f_{3L})^c
\nonumber\\
f_{2L} \rightarrow (f_{4L})^c
\nonumber\\
\eta \rightarrow - \eta
\end{eqnarray}
where we have defined parity as the inversion of only the three familiar
space coordinates. Suppose we now assume that $f_{1L}$ and $f_{2L}$
are even and $f_{3L,4L}$ are odd under $Z_2$, then on the brane located at
$y=0$, 
the Fourier expansion of the $Z_2$ even fields involve only the cosine
modes (i.e. $cos\frac{\pi n y}{R}$) and the $Z_2$ odd fields involve only
the sine modes (i.e. $sin\frac{n\pi y}{R}$). As a result, the
fields $f_{3L,4L}$ vanish on the standard model brane. Clearly this
results
in a spectrum of fields on the observed 3-brane, which is asymmetric with
respect to the $f_{1,2}$ versus $f_{3,4}$. This leads to breakdown of CP
invariance, since under CP, $f_{1,2}\rightarrow f^c_{3,4}$
respectively. Note that if
we had chosen same boundary conditions for all $f_i$ fields, CP would
have remained an unbroken symmetry on the 3-brane at $y=0$. 

As a result of the asymmetric boundary conditions, all modes of the
fermion fields $f_i$ will appear
in the 4-dimensional Lagrangian symmetrically except for the lowest modes.
 The resulting low energy Lagrangian involving only the 
zero modes can be written as:
\begin{eqnarray}
\lambda \eta {f}^{0T}_{1L}C^{-1} f^{0}_{2L}
+ \mu {f}^{0T}_{1L}C^{-1} f^{0}_{2L} +h.c. 
\end{eqnarray}
which is clearly CP violating through the phase $Im (\lambda^* \mu)$.  (One 
can of course pick a phase convention to make $\mu$ real).

Let us assume that the $\eta$ field (the ``messenger'' field) couple to
the standard model fermions in the brane such as an electron.
To illustrate that CP is originally conserved in the sector of visible
fields in the brane but broken after the asymmetric orbifold conditions
are enforced, one can
couple $\eta$ to the electron as $i \eta \bar{e} \gamma_5 e$, note that 
$\gamma_5$ is needed here for CP symmetry.
Then, assuming that all fermions are charged, 
 one generate an edm for electron via the two loop diagram in Fig. 1 
\cite{barr} with ${f_1, f_2}$ in the loop.
But if the $(f_3, f_4)$ fermions were also allowed in the loop, as
would be the case if we were to impose the symmetric orbifold boundary 
condition for them, it will cancel the contribution of $(f_1, f_2)$ and 
would result in zero edm for electron.
\begin{center}
\begin{picture}(200,160)(0,0)
\Oval(100,80)(30,50)(0) \Photon(100,110)(100,160){5}{5}
\Text(110,150)[lc]{$\downarrow$ $\gamma$}
\DashArrowLine(20,20)(60,60){5} 
\Photon(140,60)(180,20){5}{5}
\Text(35,40)[rc]{$\eta$} 
\Text(80,45)[cc]{$f_1$}

\Line(97,53)(103,47) \Line(97,47)(103,53)

\Text(105,60)[cc]{$$} \Text(130,115)[lc]{$f_2$}
\Text(140,90)[rc]{$$} \LongArrow(135,102)(140,98)
\Text(60,90)[lc]{$$}
\LongArrow(60,98)(65,102 ) 
\LongArrow(85,51 )(80,53)  
\LongArrow(121,53 )(117,51)  
\LongArrow(160,55)(170,45) 

\Text(175,50)[lc]{$\gamma$}
\ArrowLine(0,20)(20,20) \Text(10,10)[cc]{$\ell$}
\ArrowLine(20,20)(180,20) \Text(100,10)[cc]{$\ell$}
\ArrowLine(180,20)(200,20) \Text(190,10)[cc]{$\ell$}
\end{picture}
\end{center}
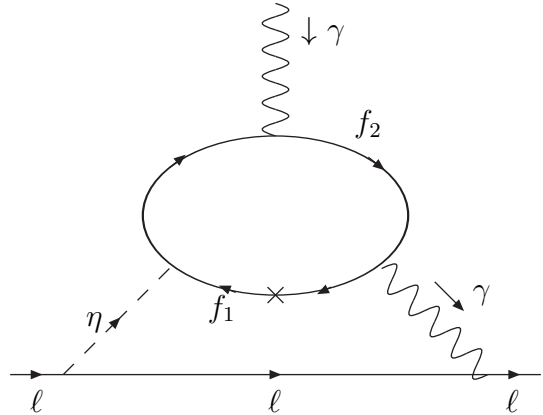
{\sl Fig.~1  The two-loop graph that contributes to edm of lepton $\ell$.  The
cross location denotes a possible mass insertion.  
In the fermion loop, if the orbifold
boundary conditions break CP symmetry, the dominant contribution comes
from the zero modes of the $f_{1,2}$ and leads to nonzero edm. On the
other hand, if the orbifold boundary conditions donot break CP, all the
KK modes are symmetric between $f_{1,2}$ and $f_{3,4}$ and due to the
negative sign in the Yukawa couplings, they cancel each other.}
\noindent

\bigskip

Note that, even without CP violating projection, at one loop level, one
can also generate operators $(f_1)^{cT}C^{-1} \sigma_{\mu\nu} \gamma_5 f_2
F^{\mu\nu}$ and $(f_3)^{cT}C^{-1} \sigma_{\mu\nu} \gamma_5 f_4 F^{\mu\nu}$
with equal coefficient (proportional to $Im (\lambda^2 \mu^*)$).  These
are usually refered to as the electric dipole moment operators of the
chiral pairs. However, their existence do not imply CP
violation.  Under CP transformation, $(f_1)^{cT} \sigma_{\mu\nu} \gamma_5
f_2 F^{\mu\nu}$ operator is mapped into $(f_3)^{cT} \sigma_{\mu\nu}
\gamma_5 f_4 F^{\mu\nu}$ and therefore CP conservation implies the two
have to have the same coefficient.  This slightly counter-intuitive
situation arises purely because the system has degenerate degrees of
freedom which makes a more general definition of CP possible. 

Note that with the coupling to the electron, the theory is CP violating
even for vanishing $Im (\lambda^* \mu)$.  This is because the coupling to
electron defines $\eta$ as CP odd, as originally imposed, in the mean
time, the coupling to $f_{1,2}$ in Eq.(3) is consistent with this CP
definition only for the special case that $\lambda$ is purely
imaginary.  Therefore for generic $\lambda$ CP is broken.

Another independent manifestation of how the orbifold condition can lead to CP
violation on the brane is to note that CP invariance for the $(\eta,
f_i)$ sector implies that if we choose a potential for the $\eta$ field
as $V(\eta) = m^2_{\eta}
\eta^2 + \lambda_{\eta}\eta^4$, with $m^2_{\eta}> 0$, then $<\eta> =
0$. Thus vacuum also leaves CP as a good symmetry. Now if we take one loop
effects with symmetric boundary conditions for all $f_i$'s, then the 
the tadpole diagrams will cancel between $f_{1,2}$ and $f_{3,4}$ keeping
the $<\eta>=0$ vev stable under radiative corrections. However once we
impose asymmetric boundary conditions
between $f_{1,2}$ and $f_{3,4}$, then there will be a nonvanishing tadpole
contribution leading
to a vev of the $\eta$ field and one will have spontaneous breakdown of CP
invariance. 

As the second example, one can choose to use bulk scalar instead of 
fermions $f_i$ to implement CP violating projection.  For example, one can 
replace $f_i$ by a pair of bulk complex charged bosons, $b_1, b_2$ which 
transform into each other under CP symmetry.  
\begin{equation}
\lambda \eta ( b_1^* b_1 - b_2^* b_2) +   m^2 ( b_1^* b_1  +  b_2^*
b_2)  + i h \eta \bar{e} \gamma_5 e
\end{equation}
where $\lambda$ and $h$ are real couplings.
Note that $b_i$ are degenerate in mass as required by CP symmetry.  Just as 
in the case of bulk fermion model, if CP were not broken, the edm of 
electron through the two loop diagram (Fig. 1) with an inner bosonic loop
would be 
zero due to the cancellation between the contributions from the two bosons 
as explicit calculation in Ref.\cite{ckp} had shown.  However, if CP is 
broken by the orbifold construction, non-vanish edm for electron results
as expected.

A common character of all the examples is that, in higher dimensions,
there is a degeneracy in the spectrum of the bulk fields such that a more
general definition of CP is possible.  The degenerate fields can be
either fermions or bosons.  
The resulting four dimensional theory on the brane can have
either soft or hard CP violation. 
While we arrange the CP symmetry in higher dimension by hand in our 
examples, in a realistic unified theory, this symmetry may arise 
automatically or accidentally because such a higher energy theory, such as 
the superstring theory, naturally has smaller number of coupling constants 
and larger symmetry which makes particle spectral degeneracy more likely 
and thus leaves room for a broader, less conventional, definition of CP 
symmetry.  It is also true that CP violating projective condition is
arranged by hand in our mechanism as in any other CP violating mechanism
in the literature, however, we believe this is the first time that it is
implemented geometrically.  Since compactification is unavoidable in
higher dimensional theory, it should not be surprising to see CP broken
in the process given that a natural mechanism exists.

We should also emphasize that, the usual common sense, initiated by Landau, 
that the CP violation is tightly related to physical complex coupling 
constants is true only for systems without spectral degeneracy.  For 
example, in our toy model in Eq.(1), there is clearly a physical complex 
coupling constant in the relative phase between $\lambda$ and $\mu$.  
However, before one imposes the CP violating projective condition, this
complex phase does not give 
rise to CP violation.  Spectral degeneracy makes it possible to define a CP 
symmetry even though the Lagrangian contains physical complex couplings.

\section{Towards a realitsic example}

In this section, we show how the idea of
the previous section can be used to generate a
multi-Higgs\cite{weinberg}model of CP violation. 
Although this model is presently very highly constrained
by experiments, because it predicts too large a value for the neutron 
electric dipole moment\cite{desh}, $b\rightarrow s +\gamma$ etc., 
we choose to discuss this since it provides a very straight forward way to
illustrate how our idea can generate a realistic model of CP violation.

As before, consider four fermion fields $f_{1,2}$ and $f_{3,4}$ in the
bulk and a singlet field $\eta$ in the brane. Under CP, we assume
the same transformation rules as before and as we saw once the orbifold
conditions asymmetrize the fermion spectrum, the $\eta$ field acquires a 
nonzero vev. The question now is how does one transmit it to the standard model.

For this purpose,
let us now assume a multi-Higgs extension of the standard model (gauge
fields, fermions as well as Higgses) living in the brane. We assume the
model to be CP invariant under the usual definition of CP transformation
of all the fields. We couple the $\eta$ field to the brane fields in a way
that preserves CP invariance. For instance, for a three
Higgs doublet model, one can choose, the following renormalizable Higgs
potential:
\begin{eqnarray}
V(\phi_a, \eta) = V_0(\phi^{\dagger}_a\phi_a) + \sum_{a,b}
\mu^2_{ab}(\phi_a^{\dagger}\phi_b + \phi^{\dagger}_b\phi_a)\nonumber\\+
i\mu'_{ab}\eta (\phi_a^{\dagger}\phi_b - \phi^{\dagger}_b\phi_a)
\nonumber\\
+\lambda_{abcd}\phi_a^{\dagger}\phi_b\phi^{\dagger}_c\phi_d,
\end{eqnarray}
with appropriate discete symmetry to suppress flavor changing neutral current. 
Note now that once $\eta$ acquires a vacuum expectation value (vev), there
is a CP phase in the Higgs sector that will lead to complex vev for the
Higgs fields and hence to the well known Weinberg profile for CP
violation.  We shall leave the discussion of other ways of generating a 
realistic model to a future publication.

\section{Profile of geometric CP violation in MSSM}

In this section, we apply this new mechanism to generate CP violation in
the minimal supersymmetric standard model (MSSM) to provide an example of
how it can be implemented in other realistic models.
For this purpose, we start with the usual MSSM field content in the brane
(i.e. $SU(2)_L\times
U(1)_Y$ gauge group and superfields $Q, L, u^c,d^c,e^c, H_u,
H_d$) augmentd by the inclusion of a single superfield, which will be the
``messenger'' of CP violation. In the bulk we will now have N=2
supersymmetry. We will have two N=2 hypermultiplets in the brane, denoted
by its N=1 components $(H_1, H^c_1; H_2, H^c_2)$. Under CP symmetry, we
assume the MSSM fields to transform as usual i.e. $Q\rightarrow Q^*$, etc. 
The rest of the fields transform as follows:
\begin{eqnarray}
\eta\rightarrow -\eta^* \nonumber\\
H_1\rightarrow H^{c*}_2\nonumber\\
H_2\rightarrow H^{c*}_1
\end{eqnarray}
We assume the theory prior to compactification to be CP symmetric so that
the only phase in the theory is in the coupling of the $\eta$ fields to
the bulk fields $H_{1,2}$:
\begin{eqnarray}
W_{\eta}= \eta (\lambda H_1H_2-\lambda^*H^c_1H^c_2) + M_1\eta^2
\nonumber\\
+M_2(H_1H_2 + H^c_1H^c_2)
\end{eqnarray}
where $M_{1,2}$ are masses expected to be of order of the fundamental
scale of the theory. Now note that since the bulk kinetic energy leads to
a term of the form\cite{hall} $H\partial_yH^c$, the required condition for
CP violation i.e. $H$ and $H^c$ have opposite $Z_2$ parity is
automatically satisfied and CP violation in the brane will ensue rather
naturally due to asymmetric spectrum of the bulk fields.

To see the profile of CP violation, let us write down the superpotential
in the brane involving the $\eta$ fields (the usual MSSM superpotential
terms involving the MSSM fields are omitted for simplicity). To
incorporate supersymmetry breaking, we have the usual hidden sector
mechanisms in mind. We will use a singlet field $S$ to implement the susy
breaking by choosing $<F_S>= M^2 \approx (10^{11})^2$ GeV$^2$.
 \begin{eqnarray}
 W_{brane}=(i\eta + M_{wk})(a + b \frac{S}{M_{P\ell}}) H_uH_d
\end{eqnarray}
We have not written terms that are suppressed by higher powers of
$M_{P\ell}$ since their effect on CP violation is negligible.

CP violation in the MSSM arises when the field $\eta$ acquires a nonzero
vev via the tadpole diagrams involving $H_{1,2}$ fields. In the
supersymmetric limit, due to the nonrenormalization theorem of
supersymmetry, $<\eta>=0$. It is then easy to see that $<\eta>\simeq
\frac{M_{susy}}{16\pi^2}$, where $M_{susy}$ is the usual scale of
superpartner masses. This leads to a profile of MSSM CP violation
where the only CP violating terms are the $\mu$ and the $B\mu$ terms.
Furthermore, the CP phase can be naturally of order $10^{-2}$ due to the
presence of the factor $16\pi^2$ above. There is no CP phase of the usual
KM type. The detailed phenomenological implications of this model will be
the subject of a future publication.  However, we shall emphasize that
this model only provides an example of how the mechanism can provide
interesting CP violating physics.  It is in no way unique.

\section{Conclusion}

In this brief note we have pointed out a novel mechanism for breaking CP
symmetry in a geometric way using the extra compact dimensions.  The
essential idea is that the asymmetrization of the spectrum by orbifold
conditions can lead to CP violating effects. Note that this is very
different from many recent papers\cite{recent} on CP violation in models
with extra dimensions in which CP violation is put into either a Higgs
VEV on some other brane or a susy breaking VEV's.  In our case, the
mechanism is genuinely geometrical in nature.  We have also illustrated
how this mechanism can be used to generate realistic models of CP
violation in the standard model as well as MSSM. In a future theory
of everything, such as the string
theory, it is likely that the theory will be so constrained that it leaves
no room for CP violation at the fundamental level.  In that case it is
interesting to entertain the idea that CP violation arises out of the
``twisting and turning'' of the compactified extra
space en route to producing the four dimensional world we live in.

D. C. wishes to thank W-Y. Keung, L. Wolfenstein, I. Low and W.-F. Chang
for discussions and Theory 
Group of Univ. Maryland for hospitality while this work was developed. 
The work of R. N. M. is supported by the NSF grant No. PHY-9802551 and that 
of D. C. supported in parts by National Science Council of R.O.C. and by 
U.S. Department of Energy (Grant No. DE-FG02-84ER40173).

\end{document}